# Tensor denoising of high-dimensional MRI data

Jonas L. Olesen[1, 2], Andrada Ianus[3], Leif Østergaard[1], Noam Shemesh[3], Sune N. Jespersen[1, 2,*]

1. Center of Functionally Integrative Neuroscience (CFIN) and MINDLab, Department of Clinical Medicine, Aarhus University, Aarhus, Denmark
2. Department of Physics and Astronomy, Aarhus University, Aarhus, Denmark
3. Champalimaud Research, Champalimaud Foundation, Lisbon, Portugal

*Corresponding author. CFIN/MindLab and Dept. of Physics and Astronomy, Aarhus University, Universitetsbyen 3, building 1710, 8000 Aarhus C, Denmark.

E-mail address: sune@cfin.au.dk

## Abstract

The signal to noise ratio (SNR) fundamentally limits the information accessible by magnetic resonance imaging (MRI). This limitation has been addressed by a host of denoising techniques, recently including so-called MPPCA: Principal component analysis (PCA) of the signal followed by automated rank estimation, exploiting the Marchenko-Pastur (MP) distribution of noise singular values. Operating on matrices comprised by data-patches, this popular approach objectively identifies noise components and, ideally, allows noise to be removed without introducing artifacts such as image blurring or non-local averaging. The MPPCA rank estimation, however, relies on a large number of noise singular values relative to the number of signal components to avoid such ill effects. This condition is unlikely to be met when data-patches and therefore matrices are small, for example due to spatially varying noise.

Here, we introduce tensor MPPCA (tMPPCA) for the purpose of denoising multidimensional data, for example from multi-contrast acquisitions. Rather than combining dimensions in matrices, tMPPCA utilizes each dimension of the multidimensional data's inherent tensor-structure to better characterize noise, and to recursively estimate signal components. Relative to matrix-based MPPCA, tMPPCA requires no additional assumptions, and comparing the two in a numerical phantom and a multi-TE diffusion MRI dataset, tMPPCA dramatically improves denoising performance. This is particularly true for small data-patches, which we believe will improve denoising in cases of spatially varying noise.

# 1. Introduction

Due to its non-invasive nature, Magnetic Resonance Imaging (MRI) is widely used in diagnostic imaging as well as in biomedical research. The Signal to Noise Ratio (SNR) of MRI is inherently low, however, meaning that image quality (spatial resolution, SNR) often competes against the time research subjects, and particularly vulnerable patients, can participate in an MRI session. While this applies to all MRI methods, some are more affected by SNR limitations as they require active signal attenuation to produce contrast. For example, diffusion MRI (dMRI) (Jones, 2010a) holds great promise as a technique for interrogating neural microstructure (Novikov et al., 2019), but it is especially limited by noise since the signal is attenuated by the application of diffusion gradients for image contrast, and by relaxation effects (Jones, 2010b). Noise thus constitutes a major limitation for inference whether by qualitative visual inspection or quantitative modelling, and a source that cannot be expected to be solved by hardware innovation alone as current hardware already achieves ≈ 90% of the Ultimate Intrinsic SNR (Ocali and Atalar, 1998; Lattanzi and Sodickson, 2012; Fan et al., 2016). In addition, note that increasing the signal-to-noise either via increases in field or by using cryogenic coils (Baltes et al., 2009) come at a relatively high cost and may still benefit from denoising to further extend their reach.

Denoising can partly alleviate this SNR limitation and is therefore an important step in the analysis of MRI data. Accordingly, various denoising methods have been adopted within the MRI community, notably non-local means filtering (Coupé et al., 2008; Manjón et al., 2012; Chen et al., 2016), total variation minimization (Knoll et al., 2011), discrete cosine transform filtering (Manjón et al., 2012), and local principal component analysis (PCA) (Manjón et al., 2013; Veraart et al., 2016b). Local PCA divides the MRI dataset into local patches, each of which is projected into a low-rank approximation given by the most significant principal components. The MRI signal is typically well described by a few such components (Veraart et al., 2016a), while thermal noise will be evenly distributed among all the components, thereby enabling noise reduction in proportion to the square root of the number of discarded components. The number of significant signal components can be estimated objectively by exploiting that noise singular values are distributed according to the Marchenko-Pastur (MP) distribution (Marčenko and Pastur, 1967). While being a seasoned idea within MRI as such (Sengupta and Mitra, 1999; Ding et al., 2010), it was only recently introduced to dMRI and coupled with local PCA as MPPCA (Veraart et al., 2016a, 2016b). The latter can be seen as a more "objective" approach to denoising, easing its application relative to earlier approaches that chose or empirically estimated the number of signal components in a more *ad hoc* manner. In particular, MPPCA was shown to compare favorably to other state-of-the-art denoising techniques in terms of noise reduction without compromising anatomical detail or otherwise blurring the data (Veraart et al., 2016b).

Because of its objective rank estimation and effectiveness, MPPCA has become a very popular SNR-boosting technique (Zhao et al., 2018; Cordero-Grande et al., 2019; Does et al., 2019; Lemberskiy et al., 2019; Ades-Aron et al., 2020; Ma et al., 2020; Ianuş et al., 2021). The implementation introduced by (Veraart et al., 2016a, 2016b) shapes the data within each patch into a voxel × modality matrix. The current formulation of the objective MP rank estimation relies on the number of signal components being small relative to the matrix dimensions, since the employed distribution of noise singular values only formally applies in the limit of infinitely large matrices and absence of signal components. This is problematic in practical applications where the denoised matrix is not very large, such as scenarios with small patches for instance due to spatially varying noise. Here, we present a revised distribution, which is applicable even when the number of signal components cannot be considered small. We thereby extend the applicability of MPPCA.

Building on the success of MPPCA, we further propose a generalization to multidimensional data, for instance, arising with multi-contrast acquisitions. The structure of such datasets is not utilized by matrix/standard MPPCA, which partially discards the information in data structure by reshaping it into a matrix. We thus propose tensor MPPCA (tMPPCA), which retains the patches' natural tensor-structure and improves the noise reduction by utilizing the additional redundancy available in the extra dimensions. To achieve this, tMPPCA builds upon higher-order singular value decomposition (HOSVD), which has previously been employed in various forms for denoising dMRI and MRI more generally (Zhang et al., 2015, 2017; Brender et al., 2019; Chen et al., 2020; Wang et al., 2020; Kim et al., 2021). Since HOSVD comprises a sequence of regular SVDs, we can exploit that objective MP rank estimation is applicable to each dimension separately, thereby extending the application of MPPCA to multidimensional data. The denoising performance of tMPPCA is further enhanced relative to regular HOSVD approaches by recursive estimation of the signal components – i.e., reducing the noise for one dimension improves the signal estimation for the other dimensions.

Here, we compare the performances of tMPPCA and MPPCA in multi-echo diffusion data as an example scenario of multidimensional data, and in simulations using a numerical phantom. Tensor MPPCA performs especially well relative to MPPCA for small patch sizes, which is relevant in the case of spatially varying noise.

## 2. Theory

The objective rank estimation in MPPCA is facilitated by the MP distribution of noise singular values as detailed in (Veraart et al., 2016b). Here, we reiterate the approach and modify the distribution of noise singular values to take the effect of a non-zero number of signal components into account. In essence, the approach considers an $M$ x $N$ matrix $X$ with singular value decomposition (SVD)

$$X = USV^\dagger \qquad (1)$$

where $U$ and $V$ are unitary matrices and $S$ is a diagonal matrix containing the singular values. Let the $P$ largest singular values represent signal components while the smaller singular values are due to zero mean independent and identically distributed (iid) noise. The squared noise singular values approximately follow the MP distribution (rescaled here to account for missing $1/\sqrt{N}$), which has a hard upper bound dictated solely by the noise variance $\sigma^2$ and $X$'s dimensions. Approximating $\sigma^2$ (details below) thus determines the upper bound and consequently the number of signal and noise components. The idea is illustrated in Fig. 1.

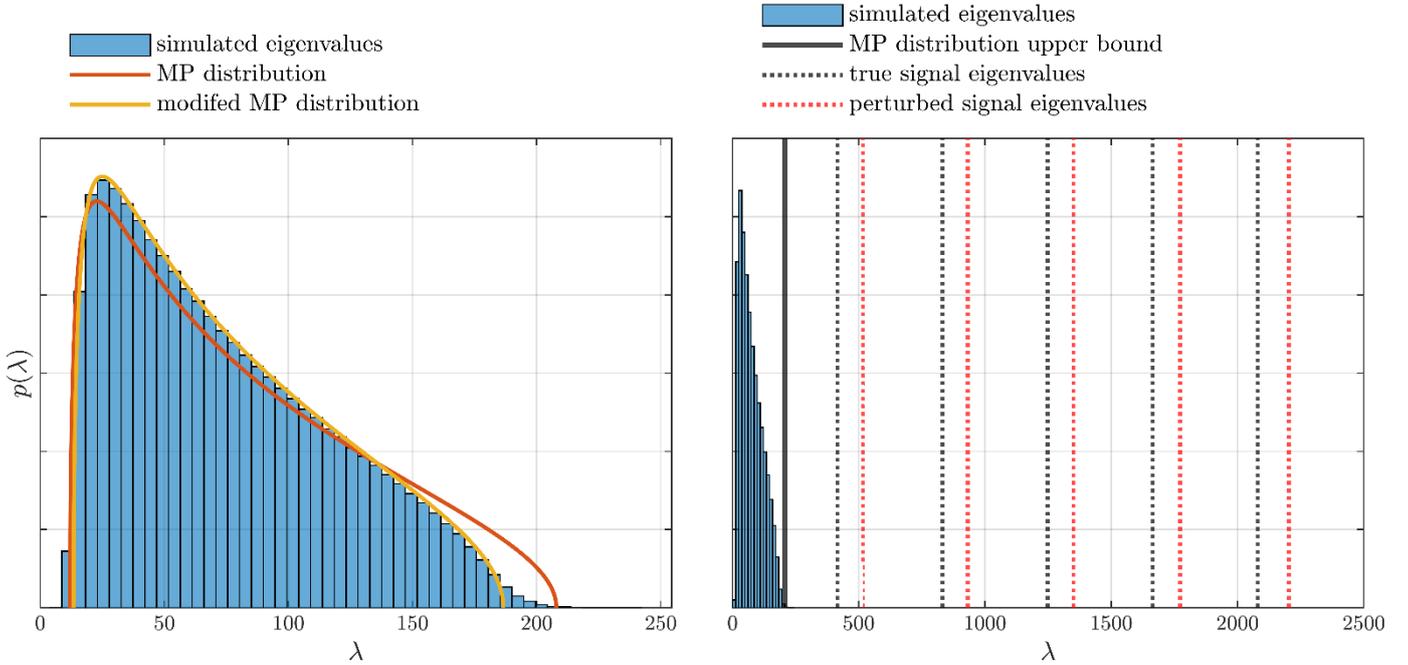

Fig. 1. Distribution of squared singular values of a $M$ = 30, $N$ = 80 matrix with $\sigma^2$ = 1 Gaussian noise and $P$ = 5 signal components with values given in the plots as black dotted vertical lines. Without loss of generality, the matrix was generated in the basis of the signal eigenvectors. The distribution of noise eigenvalues and mean perturbed signal eigenvalues were calculate from $10^5$ noise realizations. The left panel shows the simulated distribution compared with the MP distribution given in Eq. (2) and the MP distribution modified by subtracting $P$ from $M'$ and $N'$. In the right panel, the x-range is increased to encompass the true signal eigenvalues as well as the perturbed mean values of the corresponding simulated eigenvalues.

Consider the case $P = 0$, and define $N' \equiv \max(M, N)$ and $M' \equiv \min(M, N)$. Assuming X to be full rank due to noise, $M'$ is then the number of non-zero singular values. The non-zero singular values of $X$ and $X^\dagger$ are identical so the denoising is invariant towards conjugate transposing $X$. In the following, "singular values" will always be referring to the non-zero singular values. Each singular value $S_{ii}$ is associated with an eigenvalue $\lambda_i = S_{ii}^2$ of the covariance matrix $XX^\dagger$ (strictly speaking, the covariance matrix is $XX^\dagger/N$). The eigenvalues approximately follow the MP distribution. Here, we write the MP distribution scaled by a factor $M'$ relative to (Veraart et al., 2016b) and modified to exclude zero-valued singular values. Note that because of these changes, this is not strictly speaking the MP distribution, but we will refer to it as such

$$p(\lambda) = \frac{\sqrt{(\lambda - \lambda_-)(\lambda_+ - \lambda)}}{2\pi\sigma^2 M' \lambda} \quad (2)$$

$$\lambda_\pm = \sigma^2\left(\sqrt{N'} \pm \sqrt{M'}\right)^2 \quad (3)$$

Formally, the distribution only applies in the limit $M, N \to \infty$ with constant ratio $M/N$ (Marčenko and Pastur, 1967). A finite size of $X$ introduces tails in the sense that eigenvalues can exceed the otherwise hard upper and lower bounds $\lambda_\pm$ as exemplified in Fig. 1. The distribution is also affected when $P > 0$. However, the MP distribution can be heuristically generalized to non-zero $P$ simply by subtracting $P$ from $M'$ and $N'$ (also exemplified in Fig. 1). This can be verified explicitly with Monte Carlo simulations as in Fig. 1. Furthermore, Fig. 2, which shows simulations for the full range of $M/N$ and up to 80% signal ratio $P/M$, compares the eigenvalues' mean and upper bound as predicted by the modified distribution to demonstrate the applicability of the modified distribution in virtually any practical scenario.

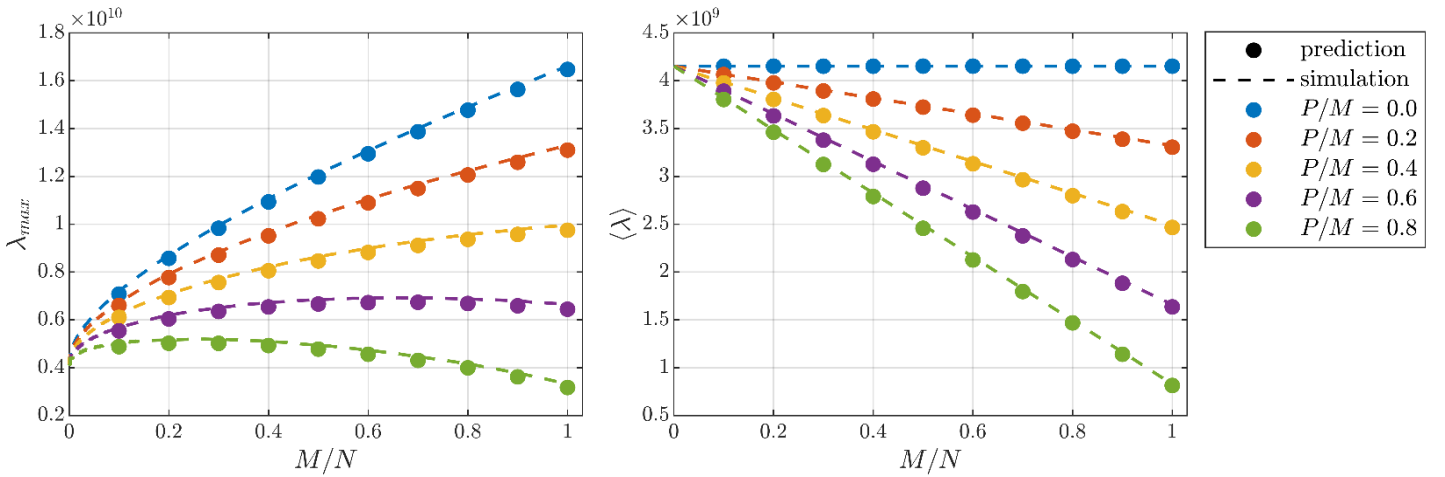

Fig. 2. Prediction and simulation of the largest and mean noise eigenvalue (squared singular value) when $X$ contains $P$ signal components. The prediction is according to the modified MP distribution (subtracting $P$ from $M'$ and $N'$ in Eq. (2)), and the simulated values were generated using $10^3$ Gaussian noise ($\sigma^2 = 1$) realizations added to a matrix with signal eigenvalues equaling 10 times the upper bound $\lambda_+$ of the MP distribution. The value of $N = 10^3$ is fixed, while the ratio $M/N$ is varied approximately over its entire relevant range from 0 to 1. Each curve is associated with a fixed value for the ratio of signal components $P/M$.

The mean of the MP distribution is $\bar{\lambda} = N'\sigma^2$. Accordingly, when taking $P \geq 0$ into account, the average sum of the noise eigenvalues is

$$\sum_i \lambda_i = (M' - P)(N' - P)\sigma^2 \qquad (4)$$

This property suggests the following algorithm for determining $\sigma^2$ and $P$ (for $\lambda_i$ sorted in descending order)

- For $P = 0, 1, \dots$
  - Calculate

  $$\sigma_P^2 = \frac{1}{(M' - P)(N' - P)} \sum_{i=P+1}^{M'} \lambda_i$$

  - Calculate

  $$\lambda_+^{(P)} = \sigma_P^2 \left(\sqrt{N'} + \sqrt{M'}\right)^2$$

  - If $\lambda_{P+1} < \lambda_+^{(P)}$, terminate and output the current value for $P$ and $\sigma^2 \approx \sigma_P^2$

In words, remove apparent signal singular values from the estimate of $\sigma^2$ until there is self-consistency – i.e. all singular values lie below the predicted MP upper bound. While the modified MP distribution yields an accurate estimate of $\sigma^2$, we found that it works best in practice to use the upper bound $\lambda_+$ from the unmodified MP distribution. This overestimates $\lambda_+$ and consequently includes less of the noise distribution's upper tail as apparent signal components (see Fig. 1).

After removing the $M' - P$ noise components, some further improvement is attainable by applying optimal shrinkage (Gavish and Donoho, 2017; Cordero-Grande et al., 2019), which approximately corrects the noise perturbation of the signal eigenvalues, illustrated in Fig. 1.

As a note of caution, we emphasize that the MP approach cannot detect signal components with variances below the upper bound of the noise distribution $\lambda_+$ and will consequently remove them. However, this is typically unproblematic in practice because the majority of the signal variance usually resides in a few large components. On the other hand, it poses a problem in scenarios with very low SNR when a non-negligible fraction of the signal variance potentially lies below $\lambda_+$.

## 2.1. MPPCA denoising

Conventional MPPCA denoising (Veraart et al., 2016a, 2016b) – which we here refer to as matrix MPCA – applies a sliding window over voxels and denoises the resulting patches independently. Each patch forms a matrix $X$ with voxels as rows and one or more modalities (i.e. diffusion, relaxation, fMRI) as columns. After calculating the SVD of $X$, the noise components are removed based on the above approach. This can be done by projecting $X$ onto the signal subspace:

$$\tilde{X} = \tilde{U}\tilde{U}^\dagger X = \tilde{U}\tilde{S}\tilde{V}^\dagger \qquad (5)$$

Here, $\tilde{U}$ is $U$ with the $M - P$ columns associated with noise eigenvalues removed. $U$ is an $M$ x $M'$ matrix and $\tilde{U}$ is an $M$ x $P$ matrix. Similarly, $\tilde{S}$ is a $P$ x $P$ diagonal matrix and $\tilde{V}$ is an $N$ x $P$ matrix.

The total noise variance in $X$ is $MN\sigma^2$, while the removed variance is given by Eq. (4). The noise variance after denoising is their difference yielding a resulting noise variance of

$$\tilde{\sigma}^2 \approx \frac{P^2 + (M-P)P + (N-P)P}{MN}\sigma^2 \qquad (6)$$

The denoised patches can be combined in one of several ways (Veraart et al., 2016b). The best performance is achieved by having a sliding stride of one voxel and averaging the patches (Katkovnik et al., 2009; Manjón et al., 2013) – that is, each voxel receives an equal contribution from each patch of which it is a member. Alternatively, one can assign only the denoised signal of the center-voxel of each patch. Notably, the remaining noise will be correlated among voxels in both cases, which is potentially problematic in some applications. If desired, this can be avoided on an image by image basis at the cost of decreased denoising efficiency by combining all voxels in one patch. This has the benefit of being fast because only one patch is processed.

## 2.2. Higher-order SVD denoising

Our proposed tMPPCA denoising is similar to denoising with HOSVD with some key differences. Here we outline the HOSVD approach to highlight these differences.

For multidimensional data, each patch has the natural structure of a tensor and $X$ can be decomposed using HOSVD. This can be done by applying regular SVD to each flattening of $X$, where flattening refers to the matrix formed by concatenating all but one index. Assume for concreteness but without loss of generality that each patch is a three-index tensor $X_{i_1 i_2 i_3}$ – for instance in dMRI with $i_1$ for voxels, $i_2$ for directions, and $i_3$ for b-values with $X$ being an $M_1$ x $M_2$ x $M_3$ tensor. The $i_1$-flattening of tensor $X$ is the $M_1$ x $M_2 M_3$ matrix $X_{i_1(i_2 i_3)}$, where (…) signifies index concatenation. In this way, there is one flattening and associated SVD for each index. Assume the number of signal components is known for each flattening. Then, a denoised tensor can be constructed analogously to Eq. (5)

$$\tilde{X} = \left( \tilde{U}_1 \tilde{U}_1^\dagger, \tilde{U}_2 \tilde{U}_2^\dagger, \tilde{U}_3 \tilde{U}_3^\dagger \right) \cdot X = \left( \tilde{U}_1, \tilde{U}_2, \tilde{U}_3 \right) \cdot \tilde{S} \qquad (7)$$

Where $\cdot$ denotes tensor multiplication in the sense that $\tilde{U}_n$ acts on the n'th index. The tensor $\tilde{S} = \left( \tilde{U}_1^\dagger, \tilde{U}_2^\dagger, \tilde{U}_3^\dagger \right) \cdot X$ with dimensions $P_1$ x $P_2$ x $P_3$ is the "core", which generalizes the diagonal matrix from SVD. For clarity, note that if $X$ is a matrix (rank-2 tensor), then the first flattening is simply $X$ while the second flattening is $X$ transposed, so $U_1 = U$ and $U_2 = V^\dagger$ in the notation above.

As with regular SVD denoising, denoising tensors with HOSVD has been performed using an *a priori* choice for the signal rank (for each dimension) (Brender et al., 2019). However, we point out that since HOSVD amounts to a sequence of regular SVD's, objective MP based rank estimation is applicable to each flattening. Thereby, MPPCA can be applied to tensors while exploiting the redundancy of the additional dimensions. For further improvement, we introduce a recursive step, which then defines the proposed tMPPCA denoising as detailed below.

## 2.3. Tensor MPPCA denoising

The proposed tMPPCA denoising modifies the HOSVD approach as follows. Consider again the example of a three-index tensor $X_{i_1 i_2 i_3}$ and take $X$'s $i_1$-flattening $X_{i_1(i_2 i_3)}$ and the associated SVD $U_1 S_1 V_1^\dagger$. Now, MP rank reduction is applicable and denoises the flattening to $\tilde{U}_1 \tilde{S}_1 \tilde{V}_1^\dagger$. Define the tensor $\tilde{X}_1 \equiv \tilde{U}_1^\dagger \cdot_{i_1} X$, where $\cdot_{i_n}$ signifies multiplication onto index $i_n$. While $X$ has dimensions $M_1$ x $M_2$ x $M_3$, $\tilde{X}_1$ has dimensions $P_1$ x $M_2$ x $M_3$, where $P_1$ is the number of identified signal components in the MP rank reduction step. Crucially, $\tilde{X}_1$ retains $X$'s noise properties because $\tilde{U}_1$ is unitary. Therefore, the procedure can be repeated by applying SVD and rank-reduction to the $i_2$-flattening of $\tilde{X}_1$ yielding $\tilde{U}_2$ and $\tilde{X}_2 \equiv \tilde{U}_2^\dagger \cdot_{i_2} \tilde{X}_1$ and so on. The recursive procedure stops at the last index after which a denoised version of $X$ can be reconstructed using Eq. (7) in the same way as for HOSVD.

Note that the SVD at step n depends on the denoising at step n-1, which improves the estimates of $\tilde{U}_n$ and $\tilde{X}_n$. In contrast, each $U_n$ is estimated independently in the HOSVD approach. Therefore, the ordering of the indices/flattenings does not matter for HOSVD but does matter for tMPPCA. Choosing the index ordering thus constitutes an additional input when applying tMPPCA in addition to the choice of sliding window. In contrast to HOSVD, optimal shrinkage is applicable to the final set of signal eigenvalues $\tilde{S}_n$, further improving performance.

Each SVD followed by MP rank reduction produces an estimate of $\sigma^2$. Since the noise properties are unchanged throughout the procedure, these estimates are ideally equal but will differ slightly due to finite dimensions and random noise. Therefore, the $\sigma^2$ estimate can be improved by combining the individual estimates in a weighted sum. As such, tMPPCA improves the noise characterization in addition to the denoising itself. However, it requires an initial pass in which the SVD of each flattening is calculated prior to any denoising.

Tensor MPPCA reduces to MPPCA when $X$ is a matrix (rank-2 tensor) thereby providing equal performance when no additional redundancy is available but improved performance otherwise. It outperforms classical HOSVD approaches due to its objective rank estimation and the recursive approach, which improves each proceeding SVD and enables the application of optimal shrinkage. An overview of the algorithm is given below assuming the indices of $X$ has been ordered according to the desired index ordering. We note that this is a conceptual outline and not an efficient implementation.

- (Calculate improved $\sigma^2$-estimate from noise singular values of all flattenings)
- Set $\tilde{X}_0 = X$
- For $n$ = 1, 2, …, $\mu$ (with $\mu$ being the number of indices in $X$, i.e. tensor order)
    - Calculate the SVD $U_n S_n V_n^\dagger$ of the $i_n$-flattening of $\tilde{X}_{n-1}$
    - Apply MP rank reduction yielding $\tilde{U}_n \tilde{S}_n \tilde{V}_n^\dagger$
    - Set $\tilde{X}_n = \tilde{U}_n^\dagger \cdot_{i_n} \tilde{X}_{n-1}$
- (Apply optimal shrinkage to $\tilde{S}_\mu$ modifying $\tilde{X}_\mu$ accordingly – the $i_\mu$-flattening of $\tilde{X}_\mu$ is $\tilde{S}_\mu \tilde{V}_\mu^\dagger$)
- Output $\tilde{X} = (\tilde{U}_1, \cdots, \tilde{U}_\mu) \cdot \tilde{X}_\mu$

Subtracting the removed variance at each step of tMPPCA from the total noise variance $M_1 \cdots M_\mu \sigma^2$ (same approach as for MPPCA) yields the average noise variance after denoising

$$\tilde{\sigma}^2 \approx \frac{P_1 \cdots P_\mu + \sum_n (M_n - P_n) P_n}{M_1 \cdots M_\mu} \sigma^2 \qquad (8)$$

## 3. Methods

### 3.1. Denoising

We provide a Matlab implementation of tMPPCA openly available at https://github.com/sunenj/MP-PCA-Denoising.

We compare the performance of matrix and tMPPCA implemented according to the theory outlined above. That is, we generalize the MPPCA implementation to use the revised MP distribution. This improves the performance in specific scenarios but will typically only result in subtle differences, and this is also the case for the example datasets used here. Using the revised MP distribution matters more for tMPPCA as this approach is more likely to encounter noisy matrices with a relatively large number of signal components because the resulting matrix after denoising in each index contains progressively less noise. In practice, we can then use the tMPPCA implementation for both tensor and MPPCA simply by only denoising using the voxel indices in the case of MPPCA.

For tMPPCA specifically, we always choose the index ordering by maximizing the amount of retained signal variance. This is sub-optimal in terms of noise reduction, which is generally maximized by discarding the maximal number of components. However, it is the most conservative approach in terms of minimizing the amount of discarded true signal variance. The idea is that the amount of true signal variance above the upper bound of the MP-distribution might vary between the flattenings of $X$ and can be estimated as the sum of squared signal singular values for each flattening. The signal singular values are inflated by noise as outlined above and this was corrected for using optimal shrinkage (Gavish and Donoho, 2017) in the estimates. In principle, the optimal index ordering according to this criterion can vary between different patches. However, at least for the datasets employed here, we found that the optimal ordering either did not vary between patches or did vary, but the retained signal variances were practically equal between the relevant indices. Therefore, for simplicity, the same ordering was used for all patches.

### 3.2. Data

All experiments were preapproved by the competent institutional and national authorities and were carried out in accordance with European Directive 2010/63.

To demonstrate the benefit of tMPPCA on datasets with multiple redundancy sources, we employ multi-TE diffusion data with a four-index structure (voxels x $\hat{g}$ x $b$ x TE). A large subset of the MR images were below the noise floor due to combined strong diffusion weighting and long TE, but we recover MR images with sufficient SNR for applying DKI.

The acquisition was performed on a mouse brain *ex vivo*. Following transcardial perfusion, the brain was removed, immersed in 4% Paraformaldehyde (PFA) solution (24 h), and washed in Phosphate-Buffered Saline (PBS) solution (48 h). Then, the sample was placed in a 10 mm NMR tube with Fluorinert kept at 22 °C and scanned using a 16.4 T Bruker Aeon scanner with a Micro5 probe (producing gradients up to 3000 mT/m).

The data was acquired using a diffusion weighted RARE sequence (remmiRARE from https://remmi-toolbox.github.io/) with field of view 10x8 mm$^2$, matrix size 100x80, in-plane resolution 100x100 µm$^2$, slice thickness 500 µm, and 27 slices. The data was acquired with partial Fourier PF = 1.2 in the phase direction with resultant matrix size 100x67. Other parameters were TR = 4 s, gradient pulse width δ = 3.5 ms, gradient pulse separation Δ = 5.6 ms, 20 gradient directions uniformly distributed on a hemisphere, and 6 b-values linearly distributed from 0.5 to 3 ms/µm$^2$. The number of echo times was 38 with TE linearly varied between 11 and 111 ms. The total scan time was 10 hours.

The raw k-space data was Fourier transformed and denoised with either matrix or tMPPCA using a 10 x 10 sliding window and patch averaging (other window sizes were used for the supplementary material and are specified there). Prior to denoising, a spike artifact specific to the employed dataset was partially corrected by replacing the relevant patch of k-space data with a similar patch from the corner of acquired k-space (effectively equivalent to replacing the data with white noise at the same noise level as the original data). This substantially suppresses the artifact, but subtle vertical lines are visible in some images. Due to the use of partial Fourier, the effective in-plane resolution of the data is reduced to 100x84 µm$^2$. Since the denoising was performed on the raw data, it was unnecessary to recover the full resolution for demonstrating the denoising performance, and this processing step was therefore skipped. To further reduce memory and compute intensity, only a single slice was used. Magnitude data was used in the proceeding DKI analysis.

### 3.3. Simulations

The denoising performance was also assessed using a numerical phantom. This provides the benefit of a known ground truth. The phantom was created by using the fitted DKI parameters from the denoised multi-TE diffusion dataset to generate a DKI signal for each TE. The TE values for the phantom are therefore identical to those acquired experimentally. However, only the 20 smallest TE values (up to TE = 62 ms) were used to ensure very high SNR for the DKI fit. The same b-values and number of gradient directions were used as in the original dataset except when stated otherwise. Gaussian noise of SNR $\equiv S_0/\sigma$ = 20 was added to the phantom data prior to denoising, and the denoising performance is quantified as the root-mean-squared-error (RMSE) between the denoised and ground truth data.

This performance metric reflects the denoising algorithm's ability to recover the actual ground truth rather than simply the amount of removed variance, which potentially includes lost true signal variance.

## 4. Results

### 4.1. Multi-TE diffusion data

The denoising performances of matrix and tMPPCA on the acquired multi-TE data are compared in Fig. 3. Tensor MPPCA benefits considerably from the redundancy of the multiple echo times achieving an estimated three-fold SNR gain relative to MPPCA. Since the ground truth is unknown, the SNR gains are estimated using Eqs. (4) and (8) respectively. At small TE, the denoised images are indistinguishable between matrix and tMPPCA, but at large TE, the improved quality of the images recovered using tMPPCA is apparent. Notably, in the matrix denoising, many features of the brain at long TE are lost, but the tensor denoising clearly better preserves the information in terms of agreement with expected tissue symmetries and features. The noise reduction is appreciated by comparing single voxel signals (Fig. 3 C), where the expected monotonous signal decay can be appreciated only from the tensor denoising. Figure 3 also includes residual distributions demonstrating lack of structure and excellent agreement with Gaussian statistics suggesting that both matrix and tMPPCA removes a negligible amount of true signal variance relative to the amount of removed Gaussian noise.

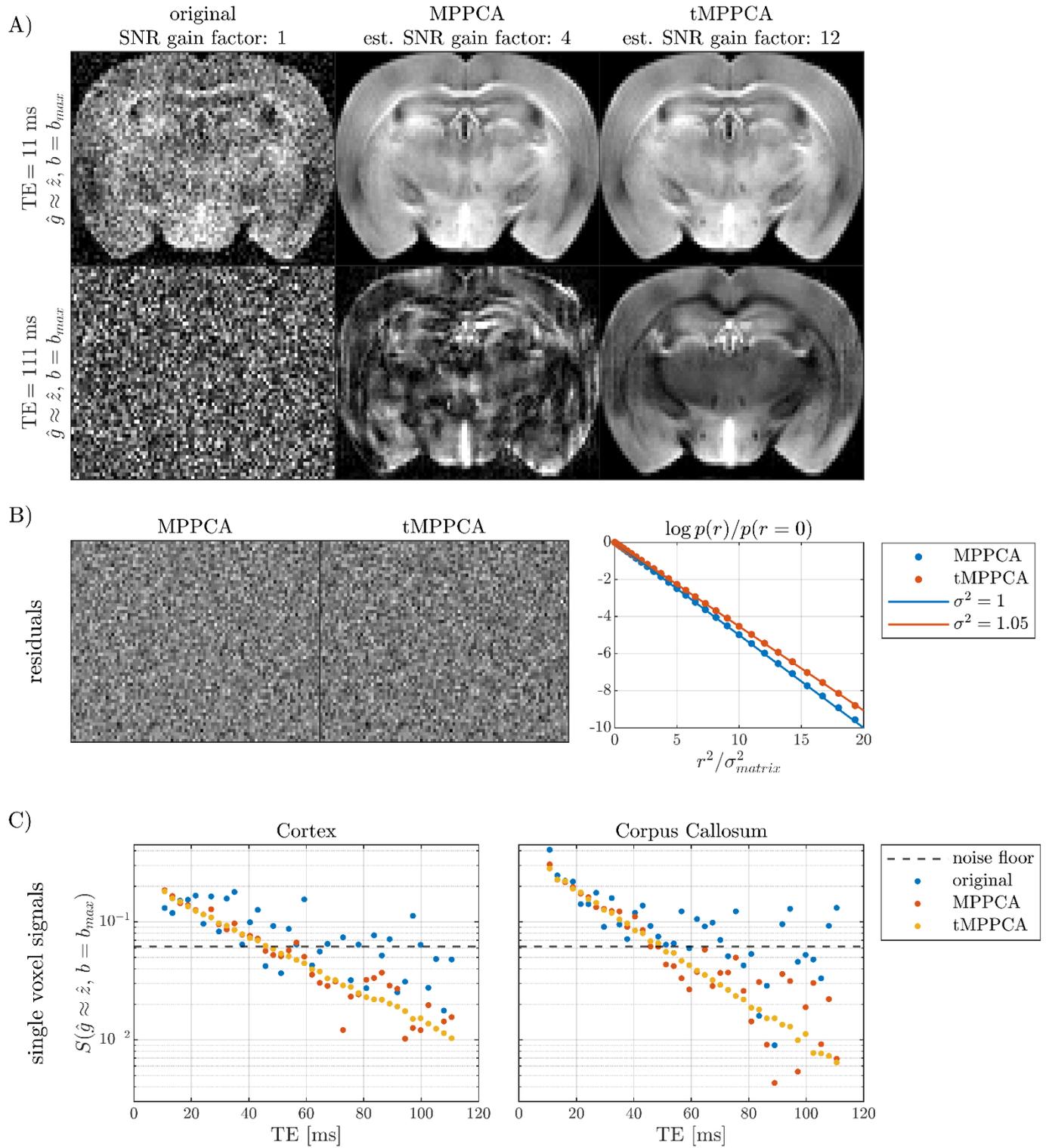

Fig. 3. Comparison of matrix and tMPPCA denoising using a 10 x 10 sliding window and index-ordering: voxels x TE x $\hat{g}$ x $b$. A) Magnitude of original and denoised image examples. The quality improvement of the recovered images from matrix to tMPPCA can be appreciated at large TE. The SNR gains are estimated using Eq. (4) and (8) respectively. B) Examples of residual images (real part) and log of the residual distributions compared to Gaussian reference lines. The variance is relative to MPPCA:

tMPPCA removes 5% ≈ $1/4^2$ - $1/12^2$ (cf. Fig. 3A) additional variance in this dataset. C) Example signals for single voxels located centrally in the Cortex and Corpus Callosum. The remaining noise is not correlated along non-voxel indices and thus immediately visible. The dashed lines indicate the Rician noise floor for the non-denoised data.

This example dataset consists of a large number of images with negligible spatial variation in the noise variance. Accordingly, the performance of MPPCA is improved by using a larger sliding window. In the limiting case of denoising all voxels as one patch, MPPCA achieves a factor of 2.5 increased SNR gain relative to the 10 x 10 window used when producing Fig. 3 (see supplementary Fig. S.1), while the performance of tMPPCA decreases and becomes comparable but still larger. MPPCA then reproduces the tissue symmetries and features otherwise only visible for tMPPCA at large TE in Fig. 3A. This provides a verification of the denoised images from tMPPCA (at least with MPPCA regarded as a standard).

In contrast, supplementary Fig. S.2 shows the contents of Fig. 3 produced using a 3 x 3 sliding window. The performances of matrix and tMPPCA are both decreased and slight deviations from Gaussian residuals is observed for MPPCA, as can be expected due to violating the large matrix assumption in the MP rank reduction. TMPPCA is less affected since it can rely on redundancy in the other indices and maintains good performance despite the small window size.

Fig. 4 compares DKI parameter maps for the subset of denoised data with smallest and largest TE. Again, the maps show little differences at the smallest TE where the SNR is high, but at the largest TE, they are distorted substantially when using data denoised with MPPCA. This is mediated to a large degree by using the data denoised with tMPPCA.

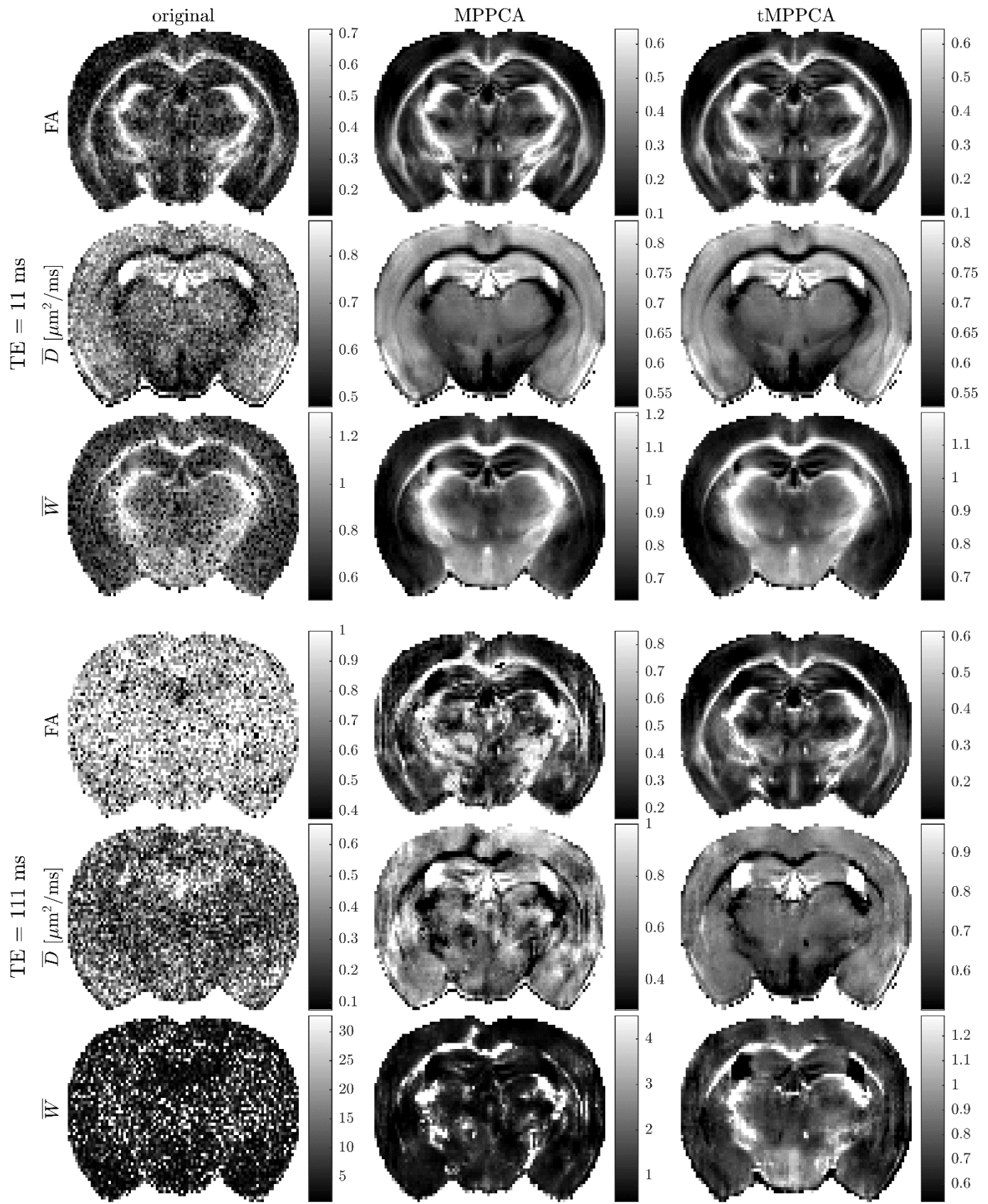

Fig. 4. Comparison of diffusion parameter maps calculated using the subset of data with smallest/largest TE and denoised with matrix and tMPPCA as labelled in the figure.

## 4.2. Simulations

Fig. 5 compares the performance of matrix and tMPPCA applied to simulated multi-TE diffusion data with varying number of echo times, gradient directions, and varying window size and patch method. Comparing the performances at varying number of echo times and gradient directions shows that tMPPCA benefits substantially from the additional redundancy this introduces. In comparison, the performance of MPPCA plateaus early. Tensor MPPCA outperforms MPPCA for all window sizes, consistent with MPPCA being a special case.

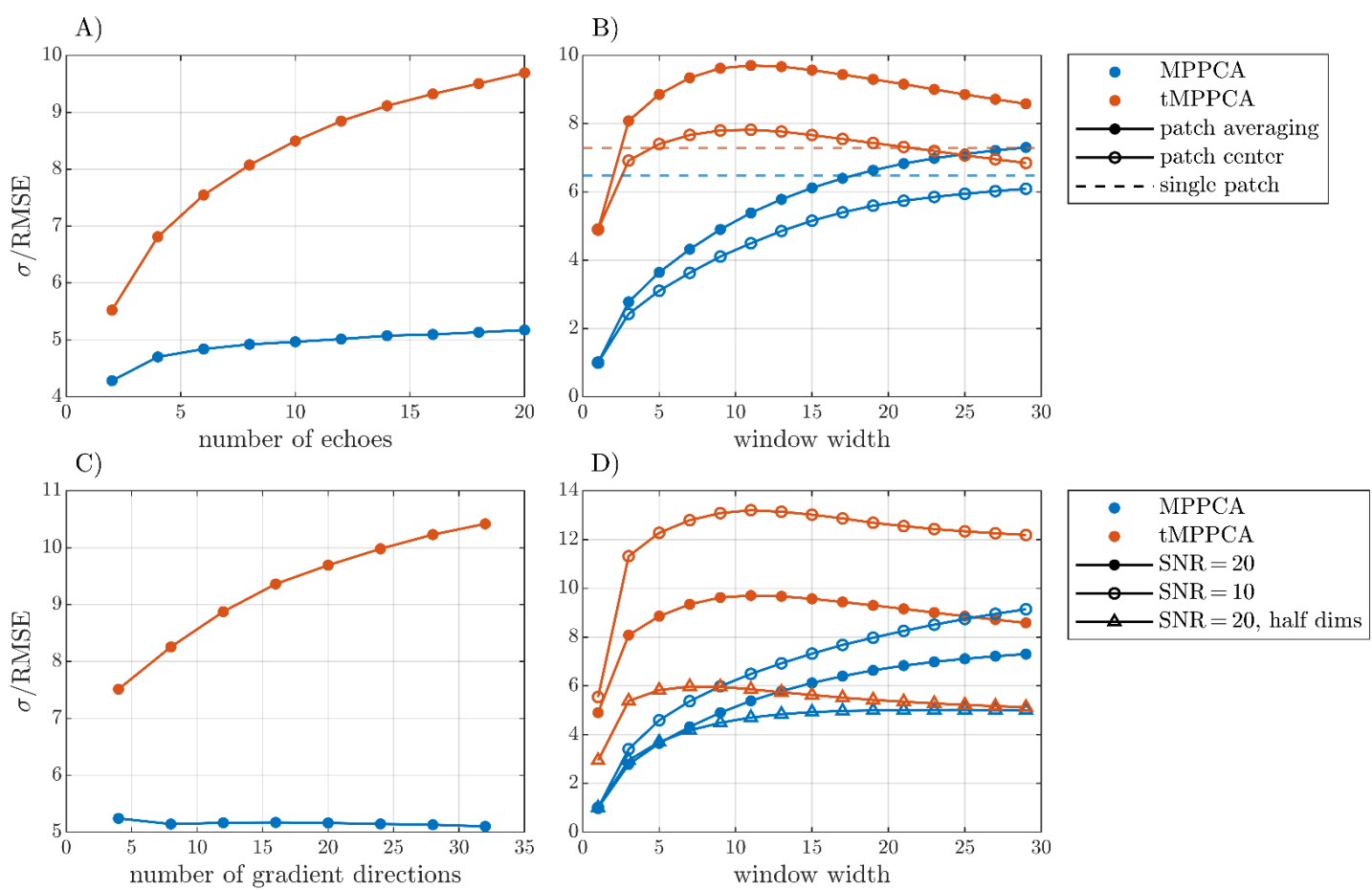

Fig. 5. SNR gain from matrix and tMPPCA applied to a numerical phantom of multi-TE diffusion data (6 b-values from 0.5 to 3.0 mu$^2$/ms, 20 gradient directions, 20 echoes from 11 to 62 ms, and SNR = 20 unless stated otherwise). A), C) The performance as a function of the number of echoes and gradient directions, respectively, using a 10 x 10 sliding window. B) The performance for varying window size and patch combination methods. Single patch refers to denoising all voxels as one patch. For patch

averaging, the denoised signal of each voxel is an average over the contributions from all patches that includes the voxel. For patch center, only the result for the center voxel is used from each denoised patch. D) The performance for varying window size using datasets with different SNR and dimensions. "Half dims" refers to a smaller dataset with half number of b-values, gradient directions, and echoes. Generally, the index ordering was voxels x TE x $\hat{g}$ x $b$ except in the case of single patch denoising for which it was TE x $\hat{g}$ x $b$ x voxels.

Generally, the performance increases with window size until reaching an optimum, which depends on the dataset size, patch combination method, and SNR. Tensor MPPCA reaches optimal performance at considerably smaller window sizes than MPPCA, for which the optima are not contained within the considered window size interval. Nevertheless, it is implicitly seen that MPPCA also reaches an optimal window size, which is smaller than the image size, because the performance when denoising all voxels as one patch is less than at the otherwise largest considered window. The performance varies slowly with window size around and especially beyond the optimum suggesting that MPPCA generally is insensitive to the choice of window size so long as it is chosen reasonably close to or larger than the optimal size. The simulations also indicate that larger window sizes are more beneficial at low SNR and/or for large datasets and vice versa.

Comparing the patch combination methods, we reaffirm that patch averaging provides the best performance with a meaningful performance gap to the method of assigning just the center voxel of each patch. Denoising all voxels as one patch provides competitive performance for large datasets or/and low SNR in the case of MPPCA in accordance with the general dependence on window size. This is not the case for tMPPCA, since it reaches optimal performance at much smaller window size.

## 5. Discussion and conclusions

This paper introduces tMPPCA, which generalizes MPPCA (Veraart et al., 2016a, 2016b) to exploit the natural tensor-structure of multidimensional data. In essence, tMPPCA builds upon the success of MPPCA by applying it sequentially when the data structure allows it. Using multi-TE diffusion data as an example, we demonstrated substantial performance improvements in terms of SNR gain using tMPPCA relative to MPPCA. This was strongly indicated by obvious visual improvements of example MR images and signal curves without introducing residuals with spatial structure or deviations from a Gaussian residual distribution. The increased noise suppression improves the potential for dMRI quantification as exemplified by DKI, which showed improvements in terms of quality of the parameter maps. The method can be applied without significantly increased computation time and is in fact faster when compared at peak performance (see below).

The results were verified and extended using a numerical phantom. In particular, we studied the effect patch size and found that while tMPPCA outperforms MPPCA for any patch size, the performance gap is largest for small sizes. Indeed, tMPPCA was found to reach optimal performance around a 10 x 10 window, while MPPCA benefitted from windows beyond 30 x 30. The specific optimal window sizes are dependent on the data, but the results demonstrate that tMPPCA better maintains performance for small patches as was also demonstrated using the acquired multi-TE diffusion data with a window size as small as 3 x 3. This is an appreciable benefit in scenarios where smaller windows are preferable due to spatially varying noise (Veraart et al., 2016a, 2016b). Also, in cases with rapidly varying contrast, the trade-off between increasing the window size and thereby increasing the number of signal components increases the relative performance using smaller patches. The results also reaffirm the original recommendation of choosing the patch size so that the resulting matrices are approximately square (Veraart et al., 2016b), but this rule of thumb does not apply to tMPPCA.

It is also notable that using smaller patches substantially decreases the computational time required for performing the denoising, which is a considerable benefit in practice. If the initial pass over all indices to improve the $\sigma^2$ estimate is skipped, the compute time of tMPPCA will typically be comparable to that of MPPCA. This is because even though tMPPCA calculates a sequence of SVDs, the dimensionality of the data decreases for each successive SVD, which places most of the computational cost in the initial SVD. If the initial pass is included, the compute time of tMPPCA is longer than that of MPPCA for equal patch sizes. However, if tensor and MPPCA are compared at their respective optimal patch sizes, tMPPCA will be significantly faster because the cost of using larger patches outweighs that of calculating multiple SVDs. For instance, MPPCA using a 30 x 30 window was 5.7 times slower than tMPPCA using a 10 x 10 window including the additional pass over for an improved $\sigma^2$ estimate, which in turn was 2.7 times slower than MPPCA using a 10 x 10 window (these calculations were performed on a generic laptop for the numerical phantom data).

The merits of additional data and redundancy was shown by varying the number of gradient directions and echoes. Both tensor and MPPCA benefits from additional data, but this is much more pronounced for tMPPCA, which better exploits the additional redundancy. This opens several perspectives on particularly favorable applications such as multi-coil (Lemberskiy et al., 2019) or multi-contrast acquisitions (Shemesh et al., 2016; Veraart et al., 2018; Kleban et al., 2020; Lampinen et al., 2020). We also note that MRI data is inherently tensor-structured in terms of voxels with x, y, z indices. For the data considered here, however, we found that combining voxels in one index performs near optimal.

Nevertheless, it is conceivable that retaining the voxel tensor-structure could increase performance in other scenarios such as reported by (Brender et al., 2019).

We compared patch combination methods proposed by (Veraart et al., 2016a), namely using only the center voxel of each patch or averaging the contributions from all relevant patches for each voxel as well as denoising all voxels as one patch, which can be seen as the large-window limiting case of patch averaging. Consistent with earlier findings reviewed by (Katkovnik et al., 2009) and as exploited in the initial introduction of local PCA to dMRI (Manjón et al., 2013), patch averaging yields the best performance by a potentially large margin. For large datasets and/or low SNR, the single patch method can provide comparable performance especially for MPPCA, which benefits more from large window sizes, but presumably also for tMPPCA under specific circumstances. We used the same simple method of equal-weights patch averaging as proposed by (Veraart et al., 2016b), but tested also weighted averaging with weights chosen according to the estimated residual noise (not shown). Since this did not yield a meaningful performance increase on our test data, we kept to the simpler strategy. However, unequal weighting possibly increases performance in some scenarios (Katkovnik et al., 2009; Manjón et al., 2013). Another extension is combining voxels or patches based on another similarity metric than proximity (non-local PCA). For instance, (Zhao et al., 2018) recently combined MPPCA with a non-local strategy involving clustering. Tensor MPPCA could replace MPPCA in such a strategy.

Tensor MPPCA inherits the limitations of MPPCA but does not introduce new ones. These limitations comprise the assumption of iid noise within each patch and the fundamental limitation that it is impossible to fully separate the signal from the noise. Choosing sufficiently small patches can fulfill the requirement of iid noise within each patch if the noise level varies spatially. It is also common that the noise level varies in accordance with the signal since magnitude data is frequently used and has Rician distributed noise (Gudbjartsson and Patz, 1995; Coupé et al., 2010). Sometimes this can be disregarded without a severe performance impact (Veraart et al., 2016b), since the variance of the Rician noise only varies notably for very small SNR and thus will often only be relevant in some areas of some images. Alternatively, (Zhang et al., 2017) proposed transforming magnitude dMRI data to equalize the noise variance (Foi, 2011) prior to HOSVD denoising, and recently (Ma et al., 2020) generalized the approach in a framework combining noise estimation, variance stabilization, and denoising. There, MPPCA with optimal shrinkage was employed, but tMPPCA can readily be used with that framework as well. If complex data is available, perhaps the simplest approach is to apply the denoising prior to calculating the magnitude data as done in the examples here. This also potentially benefits any proceeding reconstruction steps.

That signal and noise cannot be fully separated is partially manifested in inflation of signal singular values, which can be approximately corrected by optimal shrinkage (Gavish and Donoho, 2017). Furthermore, each signal singular vector is itself randomly rotated proportionally to the noise variance and its inverse signal variance (Gavish and Donoho, 2017). Accordingly, there will similarly be residual signal variance mixed across the noise components and hence discarding noise components always entails also removing some true signal variance. Therefore, it would in principle be a misconception to think that MPPCA enables noise suppression without altering the underlying true signal. In scenarios with reasonable SNR, the true signal is almost completely unaffected, but at some point as the SNR is lowered, MPPCA will remove a significant amount of true signal variance. In practice, it is beneficial for many applications to sacrifice some signal variance to remove a much larger proportion of noise variance, but one can also imagine applications where retaining the true signal variance has sufficient priority to warrant alternative strategies. For instance, the fitting of robust models might be less sensitive to noise than to removal of relatively small amounts of true signal variance in some cases. This is related to the choice of performance metric. Here, we have characterized the denoising performance in terms of least-squared error compared to the true signal. In terms of this metric, it is always beneficial to discard components with variance below the MP upper bound (Gavish and Donoho, 2017). When applying MPPCA, one should be aware of this limitation even though it will only notably come into play in scenarios with very low SNR, in which case MPPCA may still offer considerable benefit dependent on the application.

In this paper, we have extended MPPCA to better utilize the redundancy in multidimensional data. Tensor MPPCA yields substantial performance increases for such datasets and performs well for small window sizes, which is beneficial for instance with spatially varying noise. The generalization introduces no additional assumptions and reduces to MPPCA in the special case of data without tensor-structure. As such, tMPPCA can be directly adopted as is wherever MPPCA is applied.

## Acknowledgements

The authors would like to thank Dmitry Novikov for discussions. The authors are also grateful to Prof. Mark D Does and Dr. Kevin Harkins from Vanderbilt University for the REMMI pulse sequence and its analysis tools, that were supported through grant number NIH EB019980.

SJ, LØ, and JO are supported by the Danish National Research Foundation (CFIN), and the Danish Ministry of Science, Innovation, and Education (MINDLab). Additionally, LØ and JO are supported by the VELUX Foundation (ARCADIA, grant no. 00015963). NS was supported in part by the European Research Council (ERC) (agreement No. 679058). AI is supported by "la Caixa" Foundation (ID 100010434) and European Union's Horizon 2020 research and innovation programme under the Marie Skłodowska-Curie grant agreement No 847648, fellowship code CF/BQ/PI20/11760029. The authors acknowledge the vivarium of the Champalimaud Centre for the Unknown, a facility of CONGENTO which is a research infrastructure co-financed by Lisboa Regional Operational Programme (Lisboa 2020), under the PORTUGAL 2020 Partnership Agreement through the European Regional Development Fund (ERDF) and Fundação para a Ciência e Tecnologia (Portugal), project LISBOA-01-0145-FEDER-022170.

Supplementary

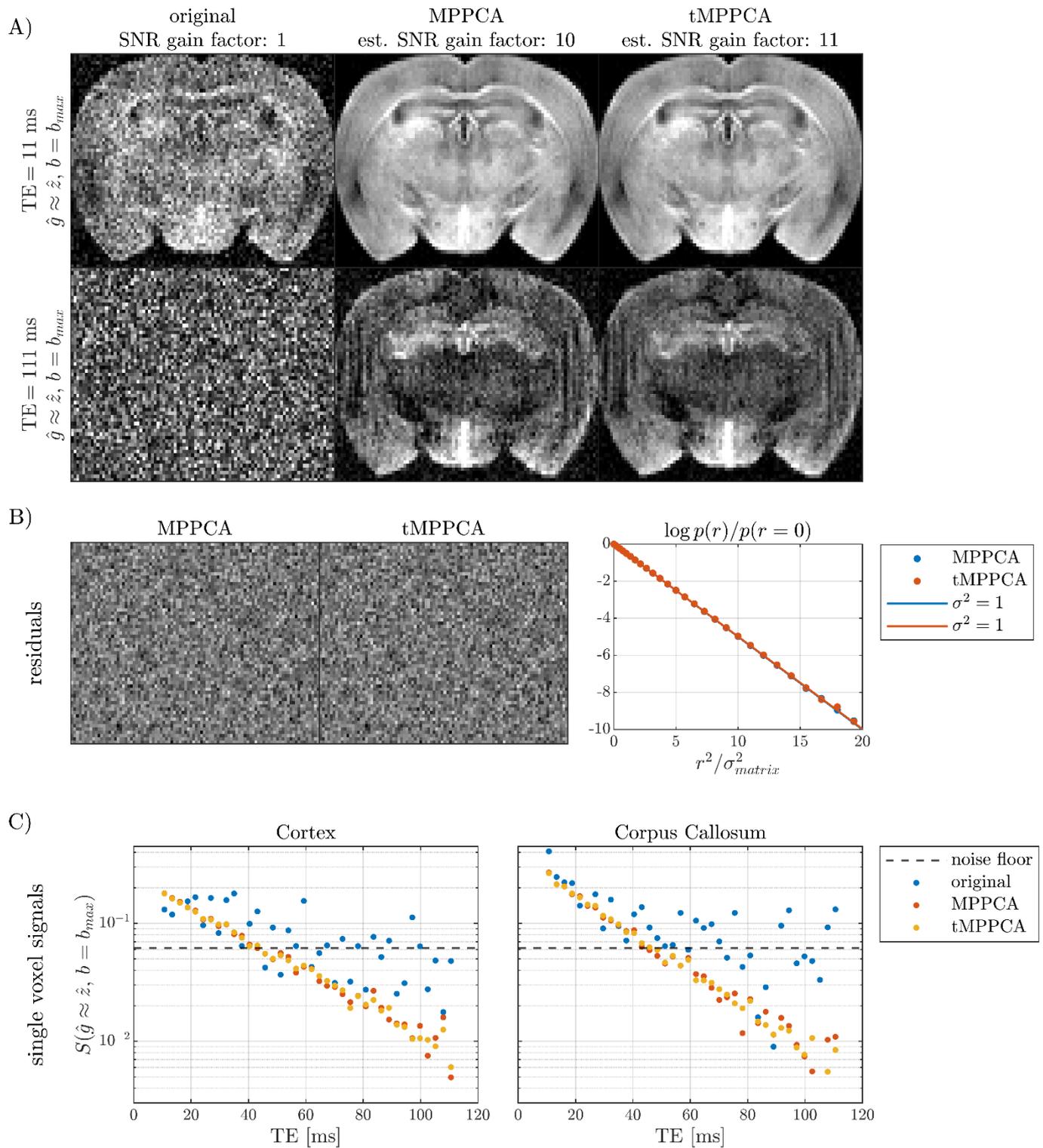

Fig. S.1. Comparison of matrix and tMPPCA denoising using a single patch including all voxels and index-ordering: voxels x ($b$, $\hat{g}$) x TE. A) Magnitude of original and denoised image examples. The SNR gains are estimated using Eq. (4) and (8) respectively. B) Examples of residual images (real part) and log of the residual distributions compared to Gaussian reference lines. C) Example signals for single voxels located centrally in the Cortex and Corpus Callosum.

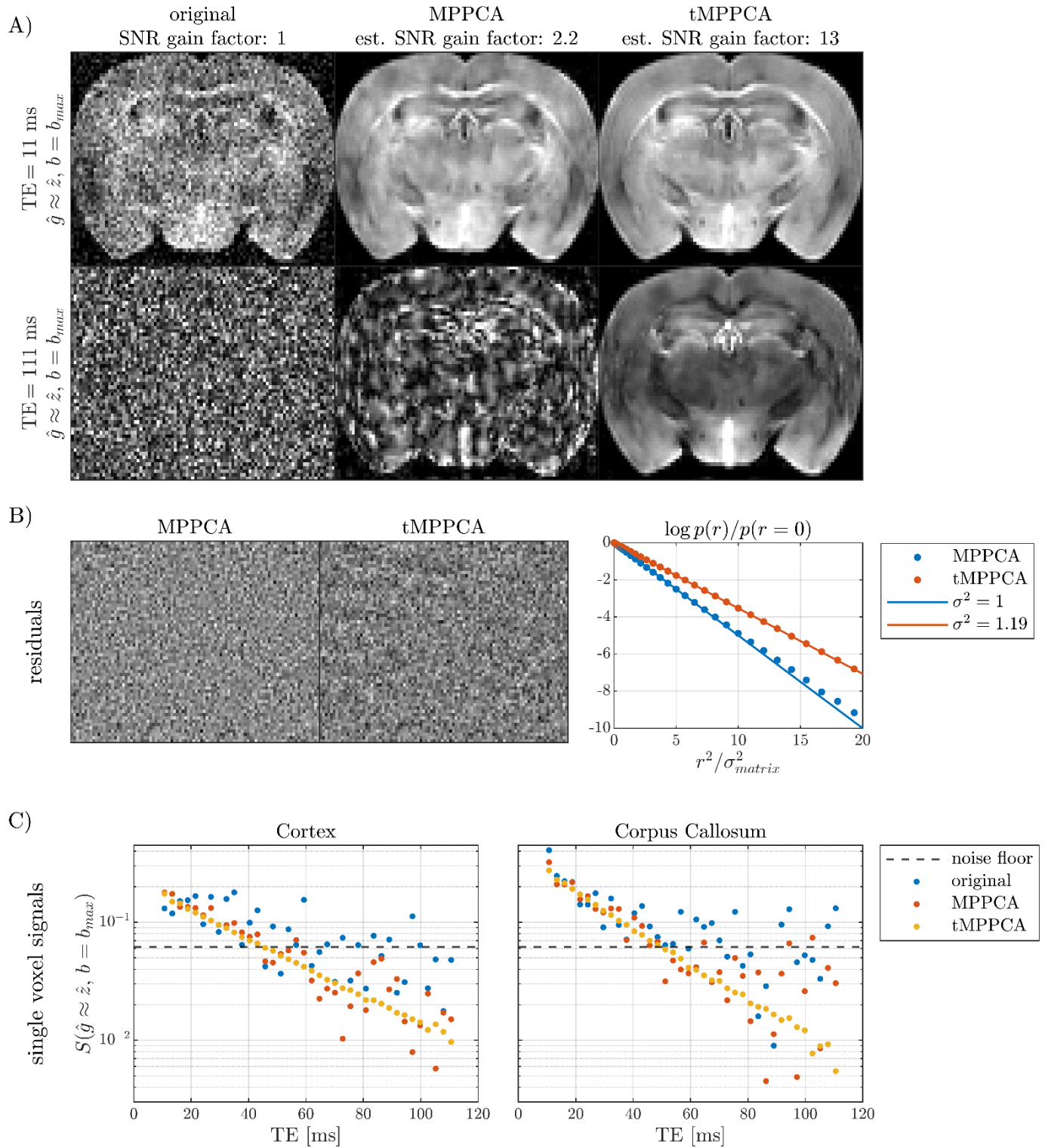

Fig. S.2. Comparison of matrix and tMPPCA denoising using a 3x3 sliding window and index-ordering TE x $\hat{g}$ x voxels x $b$. A) Magnitude of original and denoised image examples. The SNR gains are estimated using Eq. (4) and (8) respectively. B) Examples of residual images (real part) and log of the residual distributions compared to Gaussian reference lines. C) Example signals for single voxels located centrally in the Cortex and Corpus Callosum.